\documentstyle[12pt,aasms4]{article}

\def\gappr{\mathrel{\vcenter{\offinterlineskip \hbox{$>$}
    \kern 0.3ex \hbox{$\sim$}}}}
\def\lappr{\mathrel{\vcenter{\offinterlineskip \hbox{$<$}
    \kern 0.3ex \hbox{$\sim$}}}}

\def\fun#1#2{\lower3.6pt\vbox{\baselineskip0pt\lineskip.9pt
        \ialign{$\mathsurround=0pt#1\hfill##\hfil$\crcr#2\crcr\sim\crcr}}}

\def\ben{\begin{equation}}
\def\be#1{\begin{equation}\label{eq:#1}}
\def\ee{\end{equation}}

\received{}
\revised{}
\singlespace

\begin{document}

\title{Deflection of ultra high energy cosmic rays by the galactic 
magnetic field: from the sources to the detector}

\author{Gustavo A. Medina Tanco$^{1,2}$, 
Elisabete M. de Gouveia Dal Pino$^{1}$ 
and Jorge E. Horvath$^{1}$}

\affil{ 1.
Instituto Astron\^omico e Geof\'isico, University of S\~ao Paulo, 
Av. Miguel St\'efano, 4200, \\
04301-904, S\~ao Paulo, SP, Brasil}

\affil{ 2. Royal Greenwich Observatory, Madingley Rd., Cambridge CB3 0EZ,
UK, \\ 
gmt@ast.cam.ac.uk}

\singlespace

\begin{abstract} 

We report the results of 3D simulations of the 
trajectories of ultra-high
energy protons  and Fe nuclei (with
 energies $E = 4 \times 10^{19}$ and
 $ 2.5 \times 10^{20} $ eV)
propagating through the galactic magnetic field from 
the sources to the 
detector. A uniform distribution of anti-particles is backtracked
from the detector, at the Earth, to the halo of  the Galaxy.
We assume an axisymmetric, large scale spiral magnetic field 
permeating both the disc 
and the halo. 
A normal field component to the galactic plane ($B_z$) is
 also included in part of the simulations.

We find that 
the presence of a large scale galactic magnetic field 
does not generally 
affect the arrival directions of the protons, although 
the inclusion of a $B_z$ component may cause significant deflection 
of the lower energy protons ($E = 4 \times 10^{19} $ eV). 
Error boxes larger than or equal to 
$\sim 5^{\circ}$ are most expected in  this case. 

On the 
other hand, in the case of  heavy nuclei, 
the arrival direction
of the particles is strongly dependent on the coordinates 
of the particle source.
The deflection may be high enough ($> 20^{\circ}$)
as to make extremely difficult
any 
identification of the sources unless the real  magnetic field 
configuration is accurately determined. 
Moreover, not every incoming particle direction is allowed between
a given source and the detector. This generates sky patches
which are virtually 
unobservable from the Earth. 
In the particular case of the UHE events
of Yakutsk, Fly's Eye, and Akeno, they come from locations 
for which the deflection caused by the assumed 
magnetic field is not significant.

\end{abstract}

\clearpage
                              
\section{Introduction}

The detection of cosmic ray events with energies above
  $100$ EeV ($1 EeV \, = \, 10^{18} eV$) (Efimov et al. 1990,
Bird et al. 1995, Hayashida et al. 1994a), 
beyond the Greisen-Zatsepin-Kuzmin cutoff energy
(Greisen 1966, Zatsepin and Kuzmin 1966),  
 has posed a challenge for the understanding  of their origin and nature.
 Particles with energies above about $\sim 60$ EeV will 
loose substantial amounts 
of energy through interactions with the $2.7$ K cosmic microwave background
 radiation, so  that their sources must be within few tens of 
Mpc (Protheroe and Johnson 1995, Sigl, Schramm and Bhattacharjee 1994,
Medina Tanco, Gouveia Dal Pino and Horvath 1997a).

	An important clue to their origin can be 
obtained from the correlation of their arrival direction with astrophysical 
sources. 
Based on events observed with the Haverah park experiment (Watson 1994), 
Stanev et al. (1995) have suggested that the ultra high energy 
cosmic rays (UHECR) observed in the northern hemisphere  show a
 statistical preference for arrival directions close to
 the supergalactic
 plane (SGP). On the other hand, a similar 
analysis carried out with  the 
events detected  by the SUGAR experiment in the southern hemisphere 
(Kewley, Clay and Dawson 1996), and the AGASA  experiment  in Japan, support
 a more uniform distribution through out the sky, 
although 
some clustering is suggested by few groups of events in the later case 
(Hayashida et al. 1994a).
A major question that then arises about the UHECR, regards their
 propagation from the sources to the detector 
through the galactic magnetic field - 
What is the angular correlation between  the arrival directions  
of the particles and the parent sources after their 
passage through the galactic magnetic field? How much are 
they deflected in a large scale regular galactic
 magnetic field (GMF)? 

Under the assumption that the UHECR are mostly composed of  protons,
 we have examined in a previous work 
(Medina Tanco, Gouveia Dal Pino e Horvath 1997a)
the 
non-diffusive propagation of UHECR through the stochastic
 intergalactic
 and an extended  galactic halo magnetic fields, and found
 that the UHECR arriving 
in the galaxy seem  to point to the sources more strongly than 
previously believed (within error 
circles of at most $\approx 8^{\circ}$). 
In the present work, we address the propagation of the UHECR
 in the regular  large scale 
galactic magnetic field (GMF).

Stanev (1997) has recently performed a similar investigation 
by considering 
different galactic magnetic field configurations. 
Searching for a possible correlation  with sources in the supergalactic 
plane (SGP), he plotted the Volcano Ranch events above 
$2 \times 10^{19}$ eV (Linsley 1963, Egorov 1963,
Laurence, Reid and Watson 1991,
Yoshida et al. 1995),
after correcting their positions for 
deflection in the GMF, and found that some groups of events seem to 
be closer to the SGP after that correction. 
On the other hand, for a vast region of the sky 
($b \, > \,  0^{\circ}, l \, < \, 130^{\circ}$), 
he found that the corrected UHECR positions are further
 away from the SGP.

In this work, we go further in this investigation by 
performing 3 D simulations of proton and Fe nuclei trajectories  
through the GMF 
and present full-sky  maps (in galactic coordinates) of their arrival 
direction distribution in both the detector
  (after deflection in the GMF) and outside the galactic halo 
(before deflection in the GMF, where they point to the real source locations).
 Also, in order to make the visualization and data analysis easier,
 we have introduced a cube-based pixelization representation of the events
 in a similar way to the one employed in the construction of the COBE maps
 (Chan and O'Neill 1976, Smoot et al. 1992, Tegmark 1996; see below). 
With this technique, the  positions of the arrival directions 
of the Yakutsk, Fly's Eye and Akeno UHECR events 
(Efimov et al 1990, Bird et al. 1995, Hayashida et al. 1994b)
 are also plotted at the detector and outside the galactic halo. 
As we will show below, the main advantage of this procedure is to allow 
a direct and transparent identification of the source for a given magnetic 
field configuration.

\section{
The Galactic Magnetic Field model}

We  assume an axisymmetric, 
spiral field without reversals extending to galactocentric 
distances of 20 kpc in all radial directions, and with an even
 (quadrupole type) parity in the perpendicular direction (z)
 to the  galactic plane. This configuration, so called $ASS-S$, 
is described in detail by Stanev (1997) and is entirely consistent with
the observations (Beck et al. 1996, Kronberg 1994).
 The field is taken to be
 $6.4 \mu G$ at $r \, = \, 4$ kpc and constant at this value 
in the central region of the Galaxy. 
The radial dependence is consistent with field strengths inferred 
from pulsar rotation measures. 
As in Stanev (1995), we assume an exponentially decaying magnetic 
field with height from  the galactic plane. 
Two scalelengths are adopeted, 
$z_o  \, =  \, 1$ kpc for $|z| \, < \, 0.5$ kpc and $z_o  \, =  \, 4$
 kpc for $|z| \, > \, 0.5$ kpc. 
We have also performed some simulations of the UHECR trajectories 
including the presence of a constant z-component 
in the galactic magnetic field 
($B_z \, =\, 0.3  \mu $ G) 
pointing to North in the Northern galactic hemisphere and to 
the opposite direction in the Southern hemisphere. 
This component was superimposed to the large scale $ASS-S$ 
spiral field. 
There is some observational support for  this component that would 
be associated to  the existence of a galactic wind 
(e.g., Stanev 1997).

\section{
The Simulations}

We first assume that the UHECR are mostly ionized hydrogen 
atoms (protons) of extragalactic origin 
(Hillas 1984, Rachen and Biermann 1993, Stanev et al. 1995,
Medina Tanco, Gouveia Dal Pino and Horvath 1997a). 
In order to trace the particle trajectories through the GMF, 
we apply the $reversibility$  $principle$ 
(Stanev 1997, Fl\"uckiger et al. 1991, 
Bieber, Evenson and Lin 1992)
 by 
backtracking antiprotons. An antiproton 
injected at the Earth  in a certain direction will follow the same
trajectory than a proton arriving at the same direction at the Earth. 
Assuming an isotropic distribution, we inject antiprotons at different 
galactic longitude $b$ and latitude $l$ at the Earth, and follow their 
propagation through the GMF until they exit the galactic halo. In this 
way, we can determine the particle direction outside the Galaxy before its 
direction can be altered by the GMF.

Let us start discussing the lower energies. Fig. 1 depicts 
the results for 
protons of energy
 $4.0 \times 10^{19}$ eV propagating through the
large scale GMF including the described $B_z$ component. 
About $6 \times 10^{3}$ antiprotons uniformly 
distributed were injected at the 
Earth. 
Their positional orientation at the detector is represented by 
the points (or pixels) in the sky map of Fig. 1b, in galactic
 coordinates. 
They fit the arrival directions of the protons $after$ their 
passage through the  GMF. 
The map is in galactic coordinates. Fig. 1a 
shows the orientation distribution of the same protons 
at their arrival in the Galaxy, before they are  deflected in the GMF. 
Thus, in Fig. 1a, the events point to the true source locations. 
The UHECR positions are
distributed in different shades of gray according to their deflection 
angles due to passage through the GMF. We find that, except 
for particles 
coming from very
high latitudes, there is significant deflection of   their
 trajectories. 
Error boxes larger than or equal to
$\sim 5^{\circ}$ are most expected at these energies. 
In fact, only $\sim 25 \% $ 
of the particles present deflection angles smaller  than 
$5^{\circ} $ in their trajectories through the GMF.
 When the constant $B_z$ component is set to zero, we find that 
the deflection is drastically decreased ($68 \%$ of  the particles present
 deflections
smaller than
$5^{\circ}$, and $95\%$, smaller than 
$20^{\circ}$.
The presence or absence of that "wind" magnetic field component is 
therefore very important to track the trajectories of 
the primaries inside the Galaxy.

For  the higher energy protons
 (with $E = 2.5 \times 10^{25}$ eV), we 
find a strong reduction in the deflection 
angles of  the particle 
trajectories in 
both situations, that is,  with
or without the inclusion of the $B_z$ component.
 In the first case, 
$\sim 79 \% $ of the particles show deflections smaller 
than $2^{\circ}$, while
in the second, more than $90 \% $ of  the particles 
 are deflected by less than 
$2^{\circ}$, so that the effect of the GMF
 on the highest energy particles is never important, as
expected.

Dramatic changes do occur when heavy 
particles are considered. In this case, the 
UHECR would come from an extended halo (see, e.g., Hillas 1984,
Medina Tanco, Gouveia Dal Pino and Horvath 1997a). 
Applying the same scheme, we have calculated the mapping 
detector-source for Fe nuclei at an energy
$2.5 \times 10^{20}$ eV for $B_{z} = 0$.
Fig. 2 shows the  deflection angle distribution
for the Fe  nuclei.
As in Fig. 1, the UHECR positions are
distributed in different shades of gray according to their deflection 
angles due to passage through the GMF. Fig. 2a 
depicts the Fe nuclei 
at the galactic halo border pointing to the source directions
 (before deflection), and Fig. 2b depicts the particle arrival 
directions at the detector (after deflection).
We find that the large scale 
magnetic field component causes the arrival 
direction of the particles to be strongly dependent on the
coordinates ($l$, $b$) of the particle source. 
Moreover, about $21 \% $ 
of the particles present deflection angles larger than 
$30^{\circ} $ in their trajectories through the GMF,
 and this figure 
increases to $72\%$ in the lower energy case 
(see Medina Tanco, Gouveia Dal Pino and Horvath 1997b, for details).
As a general feature, we see that in some regions, the deflection may be high 
enough as to make extremely difficult any 
identification of the sources unless the true magnetic field configuration
can be determined. Only the events coming  from   the galactic  
anti-center direction show deflections smaller than $10^{\circ}$,
in the high energy case, and this correlation is drastically
reduced in the smaller energy case.  In the  particular case of
the locations of  the  observed UHE events of
Yakutsk, Fly's Eye, and Akeno 
(Efimov 1990, Bird et al. 1995, Hayashida et al. 1994b) the 
deflection is not very large.

Finally, we notice from the empty regions
 at high galactic latitudes in Fig. 2a, 
that some galactic directions are virtually unobservable 
at the Earth, so that not every incoming particle
direction will be allowed between source and detector.
The GMF acts as a  kind of giant spectral analyser and the
forbidden and allowed regions resemble the situation 
the dipolar magnetic field of the Earth creates on the low 
energy cosmic rays.

\section{
Discussion and Conclusions}
Although calculated for a particular 
magnetic field geometry, 
the results presented here indicate how significant
 can be the UHECR 
deflections in the large scale ordered GMF if  heavy nuclei are 
required to be the primaries.
As in  the Stanev work (1997), we find that the 
magnitude of the deflections of the 
UHECR in the GMF requires their arrival directions to be 
corrected before they are compared to astrophysical 
source locations. Moreover, we find that the arrival 
directions of the incoming particles are 
strongly dependent on 
the coordinates of the parent sources. For 
 some directions of the sky, 
the deflection can be so high that a source
 identification becomes 
virtually impossible, unless the 
true galactic magnetic 
field configuration can be accurately determined. This
effect is larger for decreasing energy and with 
the inclusion of
a $B_z$ component in the large scale GMF.
Nonetheless, for the magnetic field geometry assumed here,
 the observed UHE energy Fly's Eye, Akeno, and Yakutsk
 events, come from  a region of the sky for
 which the 
deflections in the particle trajectories are
not very significant.

An interesting consequence of the results  above 
(which was also pointed out by  Stanev 1995) 
regards the strong dependence  of  the magnitude
 of  an UHECR 
deflection with the viewing area of a given
 experiment. 
This dependence could lead to shifts between 
the arrival 
directions of  incoming events detected by different 
experiments in different viewing areas.

The  current number  of UHECR events so far detected
is not large enough to delineate definite
conclusions on their origin or arrival directions. Nonetheless,  
the Auger project (1995), which proposes the construction of 
two air shower
arrays in the Northern and Southern hemispheres 
in order to provide an all-sky coverage of the events, 
will not only contribute to  improve the statistics of the events, but 
most important, will probe valuable information on the large scale
 structure of the
galactic magnetic field thus making it possible to trace 
the particle trajectories 
and source directions.

Finally, we should note that the possible
existence of large scale regular
 extragalactic 
magnetic field components (e.g. Arp 1988) could provoke further
non-negligible UHECR
deflections in the Mpc scales which would, of course, increase the 
complexity of the 
picture presented here.

\bigskip
This work has been partially supported by the Brazilian 
Agencies FAPESP and CNPq.

\newpage

\noindent
{\bf Figure captions}

\bigskip

Figure 1: Distribution of protons of energy $4.0 \times 10^{19}$ eV propagating 
through the large scale GMF. $6 \times 10^3$ antiprotons uniformly 
distributed are injected at the Earth. Their positional orientation at 
the detector is represented by the points (or pixels) in the sky map (b), 
in galactic coordinates. They fit the arrival directions of the 
protons $after$ their passage through the  GMF. A constant $B_z$ 
component has been included in this simulation. (a) shows the 
distribution of the same protons at the galactic halo border, 
before they are deflected in the GMF. Different shades of gray 
are used according to the deflection angles of UHECR due to the GMF.

\bigskip

Figure 2: Deflection angle distribution for Fe  nuclei of energy 
$2.5 \times 10^{20}$ eV propagating  through the large
scale GMF (without including a $B_z$ component). 
As in Fig. 1, the UHECR positions are 
distributed in different shades of gray according to their deflection 
angles due to passage through the GMF. (a) depicts the arrival directions 
of Fe nuclei at the galactic halo border (before deflection by the GMF), 
and (b) depicts the particle arrival directions at the detector (after 
deflection by the GMF). Note that the observed UHE events Akeno, Fly's 
Eye, and Yakutsk do not reveal strong deflections from their original 
positions.


\newpage

\begin{thebibliography}{} 

\bibitem{21} Arp, H. 1988, Phys. Lett. A {\bf 129}, 135.

\bibitem{20} The Auger Collaboration 1995, Design Report.

\bibitem{17} Beck, R. {\it et al.} 1996, Ann. Rev. Astron. Astrophys. 
{\bf 34}, 155.

\bibitem{ } Bieber, J.W., Evenson, P.A., and Z. Lin, Z. 1992,
Antarctic J. {\bf 27}, 318.

\bibitem{2} Bird, D.J. {\it et al.} 1995, Astrophys. J. {\bf 441}, 144.

\bibitem{15}  Chan, F.K., and E. M. O'Neill, E.M. 1976, 
in {\it Feasibility 
Study of a Quadrilateralized Spherical Cube Earth Database, 
Computer Sciences Corp. EPRF Tech. Rep.}, 
edited by S. I. Konjaev, Mat. Zametki {\bf 25}, 629.

\bibitem{1} Efimov, N.N. {\it et al.} 1990, in {\it Astrophysical Aspects 
of the Most Energetic Cosmic Rays}, edited by M. Nagano and F. 
Takahara (World Scientific, Singapore), p. 20.

\bibitem{ } Egorov, T.A. 1963,  in {\it Tokyo Workshop for the 
Study of Extremely
High Energy Cosmic Rays}, edited by M. Nagano (ICRC, Tokyo), p. 35.

\bibitem{18}
Fl\"uckiger, E.O.
{\it et al.} 1991, in {Proc. 22nd Int. Cosmic Ray
Conf.} (Dublin), {\bf 3}, p. 648.

\bibitem{4} Greisen, K. 1966, Phys.Rev. Lett. {\bf 16}, 748.

\bibitem{3} Hayashida, N. {\it et al.} 1994a, 
Phys. Rev. Lett. {\bf 77}, 1000. 

\bibitem{16} Hayashida, N. {\it et al.} 1994b, Phys. Rev. Lett. 
{\bf 73}, 3491.

\bibitem{11} Hillas, A.M. 1984, Ann. Rev. Astron. Astrophys. {\bf 22}, 
425.

\bibitem{10} Kewley, L.J., Clay, R.W., and Dawson, B.R. 1996,
 Astropart. Phys. {\bf 5}, 69.

\bibitem{ } Kronberg, P. 1994, Rep. Prog. Phys. {\bf 57}, 325. 

\bibitem{ } Laurence, M.A., Reid, R.J.O., and A. A. Watson, A.A. 1991,
 J. Phys. G: Nucl. Part. Phys. 
{\bf 17}, 733.

\bibitem{14} Linsley, J. 1963, in {\it Proc. Int. Cosmic Ray Conf.}
 (Jaipur) {\bf 4}, 
77.

\bibitem{7} Medina Tanco, G.A., Gouveia Dal Pino, E.M, 
and Horvath, J.E. 1997a,
Astropart. Phys. {\bf 6}, 337.

\bibitem{19} Medina Tanco, G.A., Gouveia Dal Pino, E.M., 
and Horvath, J.E. 
1997b (in preparation). 

\bibitem{5} Protheroe, R.J., and Johnson, P.A. 1995, Astropart. Phys. {\bf 4}, 
253.

\bibitem{12} Rachen, J.P., and Biermann, P.L. 1993, Astron. Astrophys. 
{\bf 272}, 161.

\bibitem{6} Sigl, G., Schramm, D.N., and Bhattacharjee, P. 1994, Astropart. 
Phys. {\bf 
2}, 401.

\bibitem{ } Smoot, G.F. {\it et al.} 1992, Astrophys. J. Lett.
 {\bf 396}, L1.

\bibitem{9} Stanev, T. {\it et al.} 1995, Phys. Rev. Lett. {\bf 75}, 3056.

\bibitem{13} Stanev, T. 1997, Astrophys. J. {\bf 479}, 290.

\bibitem{ } Tegmark, M. 1996 Astrophys. J.  Lett. {\bf 470}, L81.

\bibitem{8} Watson, A.A. 1994, talk given in 1994 Snowmass 
(to be published).

\bibitem{ } Yoshida, S. {\it et al.} 1995, Astropart. Phys. {\bf 3}, 105.

\bibitem{ } Zatsepin, G.T., and 
Kuzmin, V.A. 1966, Pis'ma Zh. Eksp. Teor. Fiz. {\bf 4}, 114.

\end{thebibliography}
\end{document}